\ifdefined\pdfminorversion\pdfminorversion=7\fi
\documentclass[aps,pra,reprint,superscriptaddress,amsmath,amssymb,longbibliography,floatfix]{revtex4-2}
\usepackage{graphicx}
\usepackage{bm}
\usepackage[colorlinks=true,linkcolor=blue,citecolor=blue,urlcolor=blue]{hyperref}

\newcommand{\sech}{\operatorname{sech}}
\newcommand{\Id}{\mathbb{I}}
\newcommand{\ket}[1]{\lvert#1\rangle}
\newcommand{\bra}[1]{\langle#1\rvert}
\begin{document}
\lefthyphenmin=2 \righthyphenmin=3
\title{Classical Bound on the Fisher Information Rate of a Dephasing-Enhanced Quantum Neuron}

\author{Peng Wang}
\email{wp002005@163.com}
\affiliation{School of Computer and Artificial Intelligence, Southwest Minzu University, Chengdu 610225, China}
\affiliation{Centre for Quantum Technologies, National University of Singapore, Singapore 117543, Singapore}

\author{Yu-Xuan Zhang}
\affiliation{School of Physics, Nankai University, Tianjin 300071, China}
\affiliation{Centre for Quantum Technologies, National University of Singapore, Singapore 117543, Singapore}

\author{Hai-Tao Ding}
\affiliation{Centre for Quantum Technologies, National University of Singapore, Singapore 117543, Singapore}
\affiliation{MajuLab, CNRS-UNS-NUS-NTU International Joint Research Unit, Singapore UMI 3654, Singapore}

\author{Leong-Chuan Kwek}
\email{kwekleongchuan@nus.edu.sg}
\affiliation{Centre for Quantum Technologies, National University of Singapore, Singapore 117543, Singapore}
\affiliation{MajuLab, CNRS-UNS-NUS-NTU International Joint Research Unit, Singapore UMI 3654, Singapore}
\affiliation{National Institute of Education, Nanyang Technological University, Singapore 637616, Singapore}

\begin{abstract}
Dephasing can sharpen the activation of a dissipative quantum neuron, but a stationary response does not determine how much input information its output delivers per unit time. For a single qubit with partial-reset updates at rate $\nu$, we derive the threshold response bandwidth and the exact local Fisher information rate of the complete record from repeated projective readouts, including their backaction. Near the activation threshold, the rate obeys $\mathcal R\leq\nu(1-s)(1+s)^2/2\leq16\nu/27$, where $s$ is the retained amplitude of each update. For every fixed finite Markovian dephasing rate, optimization over the readout interval and $s$ gives the same supremum, approached at $s=1/3$ with arbitrarily rapid ideal readout. A finite measurement dead time instead selects a finite optimal interval, which we obtain by a dimensionless two-parameter optimization. The ideal bound is saturated by a classical two-state jump process: quantum coherence changes finite-cadence performance but cannot raise the optimized bound. These results separate noise-enhanced activation from the operational information throughput of a quantum neuron and provide a hardware-aware benchmark for finite-time quantum neural processing.
\end{abstract}
\maketitle

A physical neuron maps a weighted input to a measurable output. For quantum implementations, both the activation and the time required to observe it matter: a steep response is useful only if distinguishable output records can be obtained within the available operating time. Quantum perceptrons can encode inputs in amplitudes or controlled qubit dynamics~\cite{Tacchino2019,Torrontegui2019,Mangini2020}, and have been extended to feed-forward and continuous-variable networks~\cite{Tacchino2020,Beer2020,Killoran2019}, while engineered dissipation provides an alternative route to reproducible output states~\cite{Kraus2008,Diehl2008,Verstraete2009,Barreiro2011,Fiorelli2019}. In such devices, coherence, relaxation, and readout jointly determine the response.

Within quantum machine learning, these neuron models operate under constraints from quantum encoding, dynamics, and measurement~\cite{Biamonte2017,Cerezo2021}. Quantum-neural network studies have focused on physically realizable nonlinearities and architectures ~\cite{Schuld2014,Cao2017,Abbas2021,Mitarai2018}, together with expressivity and trainability ~\cite{Schuld2021,McClean2018}. A separate practical consideration is how rapidly a device can produce statistically distinguishable measurement records, which matters for both training and inference.

In our recent work~\cite{Zhang2026}, we showed that dephasing enhances the conditional response of a partial-reset quantum Boltzmann sampler, thereby enabling collective ordering. The local channel and its stationary activation derived there provide the starting point for the present work. We therefore ask whether the enhanced response also increases the information transmitted over a fixed time. Stationary Fisher information alone cannot answer this question, because successive measurements are correlated and each readout changes the subsequent quantum state. Complete measurement records are central to quantum estimation of open dynamics~\cite{Tsang2013,Gammelmark2014,Kiilerich2014,Catana2015,Albarelli2017,Yang2023}, and sequential measurements can themselves provide a metrological resource~\cite{Mentesoglu2026}. More generally, noise limits the precision attainable from dynamical quantum resources ~\cite{Escher2011,Demkowicz2012}, making it important to distinguish noise-enhanced response from genuine finite-time information gain. For a dissipative neuron, however, the backaction changes the activation being read out, so its gain and record information must be evaluated together.

Here we obtain an exact link between activation, bandwidth, and record information for a single clamped-input neuron. We derive a tight time-normalized bound for the specified channel family and readout protocol, valid at arbitrary dephasing strength. The derivation accounts for measurement-induced changes in the activation and all correlations in the measurement record, and also applies to intervals selected according to earlier outcomes. The optimal amplitude retention is $s=1/3$, rather than the limit $s\to1$ that maximizes stationary gain; moreover, the optimum is attained asymptotically by a classical jump process. The result therefore supplies a resource-normalized benchmark for deciding whether noise-enhanced activation produces more usable data, without assuming independent samples or an unchanged state between measurements. Because network-level algorithms must acquire output statistics from their constituent quantum systems, this neuron-level bound connects open-system device physics to finite-time quantum-machine-learning performance.

\begin{figure}[!t]
 \includegraphics[width=\columnwidth]{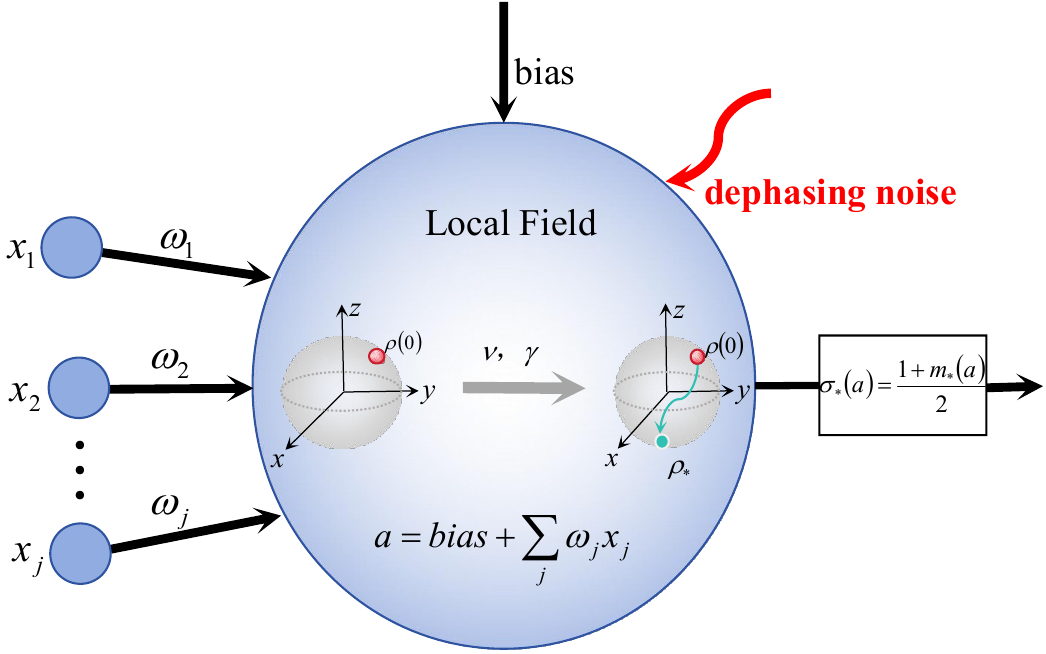}
 \caption{Schematic of the dephasing-enhanced quantum neuron studied here. The classical inputs $x_j$ and weights $\omega_j$, together with the bias, form the dimensionless local field $a=\mathit{bias}+\sum_j\omega_jx_j$. Poissonian partial-reset updates at rate $\nu$ steer the qubit from $\rho(0)$ toward the input-dependent stationary state $\rho_*$, while $Z$ dephasing acts at rate $\gamma$. A final $Z$ measurement yields the stationary polarization $m_*(a)=\mathrm{Tr}[Z\rho_*]$ and hence the stationary activation $\sigma_*(a)=[1+m_*(a)]/2$.}
 \label{fig:model}
\end{figure}

\textit{Single-neuron model.---}
The architecture is summarized in Fig.~\ref{fig:model}. The dimensionless input is $a=\mathit{bias}+\sum_j\omega_jx_j$, with externally prescribed classical inputs $x_j$, weights $\omega_j$, and a bias. Neighboring neurons are clamped; there is no self-consistency condition or many-body approximation. Write $u=\tanh a$, $q=\sech a$, and choose $Z\ket0=\ket0$. The target state is
\begin{equation}
 \ket{\psi_a}=\sqrt{\frac{1+u}{2}}\ket0+
 \sqrt{\frac{1-u}{2}}\ket1 .
 \label{eq:target}
\end{equation}
With $P_a=\ket{\psi_a}\bra{\psi_a}$ and $Q_a=\Id-P_a$, the update of Ref.~\cite{Zhang2026} is
\begin{equation}
 A_0=P_a+sQ_a,\qquad
 A_1=\sqrt{1-s^2}\ket{\psi_a}\bra{\psi_a^\perp},
 \label{eq:kraus}
\end{equation}
where $0\leq s<1$ and $\ket{\psi_a^\perp}$ is orthogonal to $\ket{\psi_a}$. The parameter $s$ retains amplitude in the orthogonal state; it is not a Hamiltonian coupling. Completeness, $\sum_\mu A_\mu^\dagger A_\mu=\Id$, defines the channel $\Phi_a(\rho)=\sum_\mu A_\mu\rho A_\mu^\dagger$. Poisson updates at rate $\nu>0$ coexist with $Z$ dephasing at rate $\gamma\geq0$:
\begin{equation}
 \dot\rho=\mathcal L_a\rho
 =\nu[\Phi_a(\rho)-\rho]+\frac{\gamma}{2}(Z\rho Z-\rho).
 \label{eq:master}
\end{equation}
This is a completely positive Lindblad evolution~\cite{Lindblad1976,Gorini1976}. The input programs the dissipative target; no coherent drive is assumed.

Write $\rho=(\Id+cX+yY+mZ)/2$, define $\kappa=\nu(1-s)$ and $d=\gamma/\kappa$, and use $c$ for coherence to distinguish it from the inputs. Direct projection gives
\begin{align}
 \dot c/\kappa&=(1+s)q-(1+sq^2+d)c-suqm,\nonumber\\
 \dot m/\kappa&=(1+s)u-(1+su^2)m-suqc,
 \label{eq:bloch}
\end{align}
with $\dot y=-(\kappa+\gamma)y$. Although dephasing has no direct population contribution, it changes the coherence entering the longitudinal equation. With $r=\gamma/[\nu(1-s^2)]$ and $\alpha=sr/(1+r)$, the established stationary solution is~\cite{Zhang2026}
\begin{equation}
 m_* =\frac{(1+\alpha)u}{1+\alpha u^2},\qquad
 c_* =\frac{q}{(1+r)(1+\alpha u^2)} .
 \label{eq:steady}
\end{equation}
The firing label is $o=(1+z)/2$, where $z=\pm1$ is the measured $Z$ eigenvalue; thus $o=1$ labels $\ket0$ and $\sigma_*=(1+m_*)/2$. Without dephasing, $\sigma_*=(1+e^{-2a})^{-1}$ and $\rho_*=P_a$ is generally coherent. A sigmoid output therefore does not imply a diagonal state. The full reset $s=0$ also preserves sigmoid populations at any $\gamma$, despite a dephasing-dependent coherence.

\begin{figure}[!t]
 \includegraphics[width=\columnwidth]{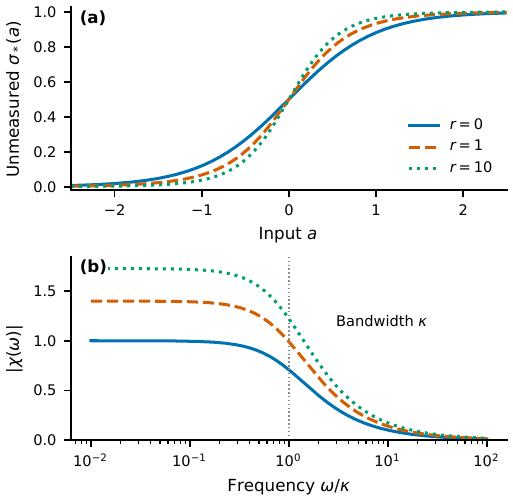}
 \caption{Stationary response before repeated readout. (a) The activation from Eq.~\eqref{eq:steady}. (b) The threshold susceptibility from Eq.~\eqref{eq:chi}. Dephasing raises the gain while leaving the relative 3-dB bandwidth $\kappa$ unchanged at fixed $s$. Here $s=0.8$ and $r=\gamma/[\nu(1-s^2)]$. All curves are theoretical.}
 \label{fig:response}
\end{figure}

\textit{Gain and bandwidth.---}
At the threshold $a=0$, the stationary coherence and longitudinal gain are
\begin{equation}
 c_0=\frac{1}{1+r},\qquad G=m_*'(0)=1+s(1-c_0).
 \label{eq:gain}
\end{equation}
For an infinitesimal modulation $a(t)=\operatorname{Re}[\epsilon e^{-i\omega t}]$ around the unmeasured steady state, the coherence has no first-order forcing. The longitudinal response obeys $\delta\dot m=-\kappa\delta m+\kappa G a(t)$, giving
\begin{equation}
 \chi(\omega)=\frac{\kappa G}{\kappa-i\omega},\qquad
 B_{3\mathrm{dB}}=\kappa.
 \label{eq:chi}
\end{equation}
The bandwidth is an angular frequency, defined relative to the zero-frequency response. Thus changing $\gamma$ at fixed $s$ enhances gain without narrowing this bandwidth [Fig.~\ref{fig:response}(b)]. Changing $s$ at fixed physical update rate is different: since $G\leq1+s$,
\begin{equation}
 B_{3\mathrm{dB}}\leq\nu(2-G),\qquad 1\leq G<2.
 \label{eq:frontier}
\end{equation}
The boundary is approached at strong dephasing for fixed $s$. Large gain near $2$ requires $s\to1$ and vanishing bandwidth. All finite-input states remain exponentially stable; the full decay matrix and spectrum are given in the Supplemental Material~\cite{Supplement}.

The stationary Bernoulli information about $a$ is $\mathcal F_Z=(\partial_a m_*)^2/(1-m_*^2)$, so $\mathcal F_Z(0)=G^2$. At the threshold this also equals the qubit quantum Fisher information~\cite{Braunstein1994,Paris2009,Toth2014,Sidhu2020,PhysRevA.105.012210,PhysRevA.109.043305,Supplement,PhysRevB.110.205402}. This is information per stationary preparation. It places no preparation-time or repeated-readout cost on the comparison, and hence is not an information rate.

\textit{Readout changes the activation.---}
Consider ideal projective measurements of $Z$ separated by an interval $\Delta>0$, with Eq.~\eqref{eq:master} acting between measurements. The input $a$ is constant but locally unknown near zero; $s,\nu,\gamma$ are known. Each outcome prepares $\ket z$ with zero transverse coherence. This protocol uses no access to the dissipative environment and no extra state preparation between readouts.

At $a=0$, coherence rebuilds after each readout as $c(t)=c_0(1-e^{-\eta t})$, where $\eta=\nu(1-s^2)+\gamma$. Writing $E=e^{-\kappa\Delta}$, the transition law near zero input is
\begin{equation}
 p(z'|z;a)=\frac{1+z'[E z+a h_\Delta]}{2}+O(a^2),
 \label{eq:transition}
\end{equation}
where
\begin{align}
 h_\Delta={}&G(1-E)
 +\kappa s c_0\frac{E-e^{-\eta\Delta}}{\eta-\kappa}.
 \label{eq:h}
\end{align}
The quotient is understood by continuity when $\eta=\kappa$. Equation~\eqref{eq:h} follows by integrating the first-order longitudinal source $\kappa[(1+s)-sc(t)]$ after projection. It accounts explicitly for the coherence removed by the measurement.

\begin{figure}[!t]
 \includegraphics[width=\columnwidth]{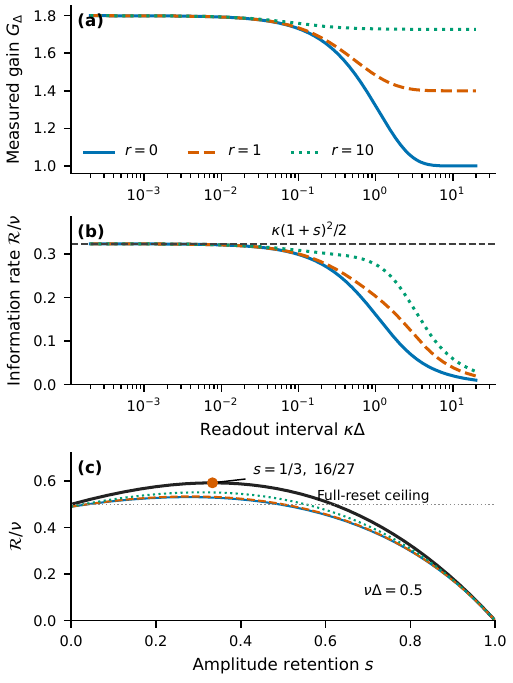}
 \caption{Information from repeated projective readouts. (a) Measurement-dependent gain and (b) exact information rate at $s=0.8$, for $r=0,1,10$. The dashed horizontal line in (b) is Eq.~\eqref{eq:bound}. (c) The bound as a function of $s$ (black), with its optimum $s=1/3$. Colored curves use $\nu\Delta=0.5$ and fixed $\gamma/\nu=0,0.36,3.6$, respectively; their $r$ values therefore vary with $s$. The dotted horizontal line is the full-reset ceiling. All rates refer to estimation of the dimensionless input at $a=0$.}
 \label{fig:information}
\end{figure}

The stationary mean of this measured process has slope $G_\Delta=h_\Delta/(1-E)$, which satisfies $G\leq G_\Delta\leq1+s$. Its limits are $G_\Delta\to G$ for $\Delta\to\infty$ and $G_\Delta\to1+s$ for $\Delta\to0$. Consequently, rapid readout sharpens the measured activation even when $\gamma=0$ [Fig.~\ref{fig:information}(a)]. Using the unmeasured activation unchanged in the likelihood would miss this backaction.

\textit{A tight information-rate bound.---}
Condition on an initial projective outcome, or start with a known $Z$ eigenstate independent of $a$. The Fisher information of a length-$n$ transition record is the expectation of its squared likelihood score. Conditional scores have zero mean, so their cross terms vanish. At $a=0$, either previous outcome yields the same information increment, $h_\Delta^2/(1-E^2)$. The exact rate for equal intervals is therefore
\begin{equation}
 \mathcal R(\Delta)=\frac{h_\Delta^2}{\Delta(1-E^2)}
 =G_\Delta^2\frac{\tanh(\kappa\Delta/2)}{\Delta}.
 \label{eq:rate}
\end{equation}
This expression retains all temporal correlations. In particular, $n$ correlated readouts cannot be replaced by $n$ independent stationary preparations.

For every $\gamma\geq0$, the rebuilt coherence is nonnegative. The driving term in the integral for $h_\Delta$ is therefore at most $\kappa(1+s)$, giving $0<h_\Delta\leq(1+s)(1-E)$. Substitution in Eq.~\eqref{eq:rate} proves
\begin{equation}
 \mathcal R(\Delta)\leq\frac{\nu}{2}(1-s)(1+s)^2\leq\frac{16\nu}{27}.
 \label{eq:bound}
\end{equation}
Here $\tanh x\leq x$ gives the first inequality, and maximizing $(1-s)(1+s)^2$ gives $s=1/3$. Both bounds are approached as $\Delta\to0$, for every fixed finite $\gamma$. The optimum is $32/27$ times the full-reset ceiling $\nu/2$, an enhancement of approximately $18.5\%$ under the same update-rate constraint.

\textit{Corollary.---}
For every fixed $0\leq\gamma<\infty$,
\begin{equation}
 \sup_{0\leq s<1,\,\Delta>0}\mathcal R_\gamma(s,\Delta)
 =\frac{16\nu}{27}.
 \label{eq:corollary}
\end{equation}
The supremum is approached at $s=1/3$ and $\Delta\downarrow0$; it is independent of $\gamma$. Thus the optimized finite-time bound is not a coherence resource: the induced classical jump process described below reaches the same value.

Real detectors impose overhead. Suppose that after each projective readout the measured eigenstate is held for a dead time $\tau_m$, during which no input-dependent evolution is accumulated. The transition likelihood is unchanged, but a cycle lasts $\Delta+\tau_m$, and hence
\begin{equation}
 \mathcal R_{\tau_m}(\Delta)
 =\frac{h_\Delta^2}{(\Delta+\tau_m)(1-E^2)}
 =\mathcal R(\Delta)\frac{\Delta}{\Delta+\tau_m}.
 \label{eq:deadtime}
\end{equation}
For every $\tau_m>0$ this rate vanishes at both $\Delta\downarrow0$ and $\Delta\to\infty$, so the best cadence is finite. The remaining optimization depends only on $\gamma/\nu$ and $\nu\tau_m$; its dimensionless formulation, full curves, and representative optima are given in the supplemental material~\cite{Supplement}.

The proof also applies when the next interval is chosen from earlier outcomes using a parameter-independent rule. Each conditional increment is bounded by $\kappa(1+s)^2\Delta_j/2$. For such schedules with a fixed total interrogation time $T$, the total conditional record information obeys $\mathcal F_{\mathrm{rec}}\leq16\nu T/27$~\cite{Supplement}. A separately informative initial preparation adds its own initial Fisher information. For a locally unbiased estimator based on the conditional record, the bound implies $\operatorname{Var}(\hat a)\geq27/(16\nu T)$.

\textit{Classical saturation and scope.---}
At arbitrarily rapid projective readout, surviving transitions are generated by the classical rates
\begin{equation}
 w(z\to-z)=\frac{\kappa}{2}(1-zu)(1-szu).
 \label{eq:classical}
\end{equation}
These are also the strong-dephasing population rates~\cite{Zhang2026}. At the threshold, the derivative of their log rate has magnitude $1+s$, while jumps occur at rate $\kappa/2$. Their trajectory Fisher information rate is precisely $\kappa(1+s)^2/2$. A classical two-state process therefore saturates the quantum finite-time bound. This is the central physical distinction: coherence can reshape a finite-cadence record, but the globally optimized ideal bound is classical. The enhancement over a full reset is a channel-design benefit and does not establish a quantum computational or metrological advantage. Projective readout does not freeze the already Markovian dissipative jumps assumed here.

The distinction is practically relevant even without invoking a network. At $s=0.8$ and $r=10$, the stationary gain is $1.727\ldots$, but the rate bound is only $0.324\nu$, below the full-reset ceiling $0.5\nu$. Increasing $\gamma$ can improve a fixed-cadence record, but cannot raise the optimized bound. The fixed resource is the attempt rate $\nu$; holding $\nu(1-s^2)$ fixed instead would define a different comparison. Ideal measurements incur no extra dead time in the bound. Finite readout cost lowers attainable throughput; weak measurements, coherent controls, and environmental monitoring require a separate optimization.

The single-neuron model therefore provides more than a modified sigmoid: it connects the activation observed under measurement to its finite-time information content. Specifically, we obtain a gain–bandwidth relation at threshold, a record likelihood that accounts for measurement backaction, and a tight bound on the information rate that reaches the classical limit. At fixed update design, dephasing can enhance the response without reducing the bandwidth, whereas optimization favors a finite retention. Extending this analysis to coupled neurons would require their joint measurement dynamics rather than simply substituting stationary activation functions.

\begin{acknowledgments}
This work is supported by the Key Research and Development Support Program of the Chengdu Municipal Science and Technology Bureau through the project ``Research, Development, and Application Demonstration Platform for an Independently Controllable Room-Temperature Quantum--Supercomputing Integrated System'' (Grant No. 2025-XT00-00012-GX). KLC acknowledges funding from the National Research Foundation, Singapore and the Ministry of Education, Singapore.
\end{acknowledgments}
\bibliography{references}

@article{Tacchino2019,
 author={Tacchino, Francesco and Macchiavello, Chiara and Gerace, Dario and Bajoni, Daniele},
 title={An artificial neuron implemented on an actual quantum processor},
 journal={npj Quantum Information}, volume={5}, pages={26}, year={2019},
 doi={10.1038/s41534-019-0140-4}}

@article{Torrontegui2019,
 author={Torrontegui, E. and Garc{\'i}a-Ripoll, J. J.},
 title={Unitary quantum perceptron as efficient universal approximator},
 journal={EPL}, volume={125}, pages={30004}, year={2019},
 doi={10.1209/0295-5075/125/30004}}

@article{Biamonte2017,
 author={Biamonte, Jacob and Wittek, Peter and Pancotti, Nicola and Rebentrost, Patrick and Wiebe, Nathan and Lloyd, Seth},
 title={Quantum machine learning},
 journal={Nature}, volume={549}, pages={195--202}, year={2017},
 doi={10.1038/nature23474}}

@article{Schuld2014,
 author={Schuld, Maria and Sinayskiy, Ilya and Petruccione, Francesco},
 title={The quest for a quantum neural network},
 journal={Quantum Information Processing}, volume={13}, pages={2567--2586}, year={2014},
 doi={10.1007/s11128-014-0809-8}}

@misc{Cao2017,
 author={Cao, Yudong and Guerreschi, Gian Giacomo and Aspuru-Guzik, Al{\'a}n},
 title={Quantum neuron: an elementary building block for machine learning on quantum computers},
 year={2017}, eprint={1711.11240}, archivePrefix={arXiv}, primaryClass={quant-ph},
 doi={10.48550/arXiv.1711.11240}}

@article{Abbas2021,
 author={Abbas, Amira and Sutter, David and Zoufal, Christa and Lucchi, Aur{\'e}lien and Figalli, Alessio and Woerner, Stefan},
 title={The power of quantum neural networks},
 journal={Nature Computational Science}, volume={1}, pages={403--409}, year={2021},
 doi={10.1038/s43588-021-00084-1}}

@article{Kraus2008,
 author={Kraus, B. and B{\"u}chler, H. P. and Diehl, S. and Kantian, A. and Micheli, A. and Zoller, P.},
 title={Preparation of entangled states by quantum {Markov} processes},
 journal={Phys. Rev. A}, volume={78}, pages={042307}, year={2008},
 doi={10.1103/PhysRevA.78.042307}}

@article{Fiorelli2019,
 author={Fiorelli, E. and Rotondo, P. and Marcuzzi, M. and Garrahan, J. P. and Lesanovsky, I.},
 title={Quantum accelerated approach to the thermal state of classical all-to-all connected spin systems with applications to pattern retrieval in the {Hopfield} neural network},
 journal={Phys. Rev. A}, volume={99}, pages={032126}, year={2019},
 doi={10.1103/PhysRevA.99.032126}}

@misc{Zhang2026,
 author={Zhang, Yu-Xuan and Chen, Jing-Ling and Kwek, Leong-Chuan and Wang, Peng},
 title={Dephasing-Enhanced Response and Phase Transitions in Quantum {Boltzmann} Samplers},
 year={2026}, eprint={2609.19591}, archivePrefix={arXiv}, primaryClass={quant-ph},
 note={Version 1}, url={https://arxiv.org/abs/2609.19591v1}}

@article{Gammelmark2014,
 author={Gammelmark, S{\o}ren and M{\o}lmer, Klaus},
 title={Fisher information and the quantum {Cram{\'e}r--Rao} sensitivity limit of continuous measurements},
 journal={Phys. Rev. Lett.}, volume={112}, pages={170401}, year={2014},
 doi={10.1103/PhysRevLett.112.170401}}

@article{Yang2023,
 author={Yang, Yaoling and Montenegro, Victor and Bayat, Abolfazl},
 title={Extractable information capacity in sequential measurements metrology},
 journal={Phys. Rev. Research}, volume={5}, pages={043273}, year={2023},
 doi={10.1103/PhysRevResearch.5.043273}}

@misc{Mentesoglu2026,
 author={Mentesoglu, Koray and Trivedi, Rahul and Mouradian, Sara},
 title={Sequential Measurements as a Resource in Ancilla-Assisted Quantum Metrology},
 year={2026}, eprint={2605.00287}, archivePrefix={arXiv}, primaryClass={quant-ph},
 note={Version 2}, doi={10.48550/arXiv.2605.00287}}

@article{Lindblad1976,
 author={Lindblad, G.},
 title={On the generators of quantum dynamical semigroups},
 journal={Commun. Math. Phys.}, volume={48}, pages={119--130}, year={1976},
 doi={10.1007/BF01608499}}

@article{Braunstein1994,
 author={Braunstein, Samuel L. and Caves, Carlton M.},
 title={Statistical distance and the geometry of quantum states},
 journal={Phys. Rev. Lett.}, volume={72}, pages={3439--3443}, year={1994},
 doi={10.1103/PhysRevLett.72.3439}}

@misc{Supplement,
 note={See Supplemental Material at [URL will be inserted by publisher] for the channel derivation, stationary state and spectrum, exact propagator, general-input Fisher information, adaptive-readout bound and finite-dead-time optimization.}}

@article{Mangini2020,
 author={Mangini, Stefano and Tacchino, Francesco and Gerace, Dario and Macchiavello, Chiara and Bajoni, Daniele},
 title={Quantum computing model of an artificial neuron with continuously valued input data},
 journal={Mach. Learn.: Sci. Technol.}, volume={1}, pages={045008}, year={2020},
 doi={10.1088/2632-2153/abaf98}}

@article{Tacchino2020,
 author={Tacchino, Francesco and Barkoutsos, Panagiotis and Macchiavello, Chiara and Tavernelli, Ivano and Gerace, Dario and Bajoni, Daniele},
 title={Quantum implementation of an artificial feed-forward neural network},
 journal={Quantum Sci. Technol.}, volume={5}, pages={044010}, year={2020},
 doi={10.1088/2058-9565/abb8e4}}

@article{Beer2020,
 author={Beer, Kerstin and Bondarenko, Dmytro and Farrelly, Terry and Osborne, Tobias J. and Salzmann, Robert and Scheiermann, Daniel and Wolf, Ramona},
 title={Training deep quantum neural networks},
 journal={Nat. Commun.}, volume={11}, pages={808}, year={2020},
 doi={10.1038/s41467-020-14454-2}}

@article{Killoran2019,
 author={Killoran, Nathan and Bromley, Thomas R. and Arrazola, Juan Miguel and Schuld, Maria and Quesada, Nicol{\'a}s and Lloyd, Seth},
 title={Continuous-variable quantum neural networks},
 journal={Phys. Rev. Research}, volume={1}, pages={033063}, year={2019},
 doi={10.1103/PhysRevResearch.1.033063}}

@article{Mitarai2018,
 author={Mitarai, Kosuke and Negoro, Makoto and Kitagawa, Masahiro and Fujii, Keisuke},
 title={Quantum circuit learning},
 journal={Phys. Rev. A}, volume={98}, pages={032309}, year={2018},
 doi={10.1103/PhysRevA.98.032309}}

@article{Cerezo2021,
 author={Cerezo, M. and Arrasmith, A. and Babbush, R. and Benjamin, S. C. and Endo, S. and Fujii, K. and McClean, J. R. and Mitarai, K. and Yuan, X. and Cincio, L. and Coles, P. J.},
 title={Variational quantum algorithms},
 journal={Nat. Rev. Phys.}, volume={3}, pages={625--644}, year={2021},
 doi={10.1038/s42254-021-00348-9}}

@article{Schuld2021,
 author={Schuld, Maria and Sweke, Ryan and Meyer, Johannes Jakob},
 title={Effect of data encoding on the expressive power of variational quantum-machine-learning models},
 journal={Phys. Rev. A}, volume={103}, pages={032430}, year={2021},
 doi={10.1103/PhysRevA.103.032430}}

@article{McClean2018,
 author={McClean, Jarrod R. and Boixo, Sergio and Smelyanskiy, Vadim N. and Babbush, Ryan and Neven, Hartmut},
 title={Barren plateaus in quantum neural network training landscapes},
 journal={Nat. Commun.}, volume={9}, pages={4812}, year={2018},
 doi={10.1038/s41467-018-07090-4}}

@article{Diehl2008,
 author={Diehl, S. and Micheli, A. and Kantian, A. and Kraus, B. and B{\"u}chler, H. P. and Zoller, P.},
 title={Quantum states and phases in driven open quantum systems with cold atoms},
 journal={Nat. Phys.}, volume={4}, pages={878--883}, year={2008},
 doi={10.1038/nphys1073}}

@article{Verstraete2009,
 author={Verstraete, Frank and Wolf, Michael M. and Cirac, J. Ignacio},
 title={Quantum computation and quantum-state engineering driven by dissipation},
 journal={Nat. Phys.}, volume={5}, pages={633--636}, year={2009},
 doi={10.1038/nphys1342}}

@article{Barreiro2011,
 author={Barreiro, Julio T. and M{\"u}ller, Markus and Schindler, Philipp and Nigg, Daniel and Monz, Thomas and Chwalla, Michael and Hennrich, Markus and Roos, Christian F. and Zoller, Peter and Blatt, Rainer},
 title={An open-system quantum simulator with trapped ions},
 journal={Nature}, volume={470}, pages={486--491}, year={2011},
 doi={10.1038/nature09801}}

@article{Gorini1976,
 author={Gorini, Vittorio and Kossakowski, Andrzej and Sudarshan, E. C. G.},
 title={Completely positive dynamical semigroups of {N}-level systems},
 journal={J. Math. Phys.}, volume={17}, pages={821--825}, year={1976},
 doi={10.1063/1.522979}}

@article{Paris2009,
 author={Paris, Matteo G. A.},
 title={Quantum estimation for quantum technology},
 journal={Int. J. Quantum Inf.}, volume={7}, pages={125--137}, year={2009},
 doi={10.1142/S0219749909004839}}

@article{Toth2014,
 author={T{\'o}th, G{\'e}za and Apellaniz, Iagoba},
 title={Quantum metrology from a quantum information science perspective},
 journal={J. Phys. A: Math. Theor.}, volume={47}, pages={424006}, year={2014},
 doi={10.1088/1751-8113/47/42/424006}}

@article{Sidhu2020,
 author={Sidhu, Jasminder S. and Kok, Pieter},
 title={Geometric perspective on quantum parameter estimation},
 journal={AVS Quantum Sci.}, volume={2}, pages={014701}, year={2020},
 doi={10.1116/1.5119961}}

@article{Escher2011,
 author={Escher, B. M. and de Matos Filho, R. L. and Davidovich, L.},
 title={General framework for estimating the ultimate precision limit in noisy quantum-enhanced metrology},
 journal={Nat. Phys.}, volume={7}, pages={406--411}, year={2011},
 doi={10.1038/nphys1958}}

@article{Demkowicz2012,
 author={Demkowicz-Dobrza{\'n}ski, Rafa{\l} and Ko{\l}ody{\'n}ski, Jan and Gu{\c{t}}{\u{a}}, M{\u{a}}d{\u{a}}lin},
 title={The elusive {Heisenberg} limit in quantum-enhanced metrology},
 journal={Nat. Commun.}, volume={3}, pages={1063}, year={2012},
 doi={10.1038/ncomms2067}}

@article{Kiilerich2014,
 author={Kiilerich, Alexander Holm and M{\o}lmer, Klaus},
 title={Estimation of atomic interaction parameters by photon counting},
 journal={Phys. Rev. A}, volume={89}, pages={052110}, year={2014},
 doi={10.1103/PhysRevA.89.052110}}

@article{Tsang2013,
 author={Tsang, Mankei},
 title={Quantum metrology with open dynamical systems},
 journal={New J. Phys.}, volume={15}, pages={073005}, year={2013},
 doi={10.1088/1367-2630/15/7/073005}}

@article{Albarelli2017,
 author={Albarelli, Francesco and Rossi, Matteo A. C. and Paris, Matteo G. A. and Genoni, Marco G.},
 title={Ultimate limits for quantum magnetometry via time-continuous measurements},
 journal={New J. Phys.}, volume={19}, pages={123011}, year={2017},
 doi={10.1088/1367-2630/aa9840}}

@article{Catana2015,
 author={Catana, C{\u{a}}t{\u{a}}lin and Bouten, Luc and Gu{\c{t}}{\u{a}}, M{\u{a}}d{\u{a}}lin},
 title={Fisher informations and local asymptotic normality for continuous-time quantum {Markov} processes},
 journal={J. Phys. A: Math. Theor.}, volume={48}, pages={365301}, year={2015},
 doi={10.1088/1751-8113/48/36/365301}}

@article{PhysRevA.105.012210,
  title = {Extracting non-Abelian quantum metric tensor and its related Chern numbers},
  author = {Ding, Hai-Tao and Zhu, Yan-Qing and He, Peng and Liu, Yu-Guo and Wang, Jian-Te and Zhang, Dan-Wei and Zhu, Shi-Liang},
  journal = {Phys. Rev. A},
  volume = {105},
  issue = {1},
  pages = {012210},
  numpages = {9},
  year = {2022},
  month = {Jan},
  publisher = {American Physical Society},
  doi = {10.1103/PhysRevA.105.012210},
  url = {https://link.aps.org/doi/10.1103/PhysRevA.105.012210}
}

@article{PhysRevA.109.043305,
  title = {Non-Abelian quantum geometric tensor in degenerate topological semimetals},
  author = {Ding, Hai-Tao and Zhang, Chang-Xiao and Liu, Jing-Xin and Wang, Jian-Te and Zhang, Dan-Wei and Zhu, Shi-Liang},
  journal = {Phys. Rev. A},
  volume = {109},
  issue = {4},
  pages = {043305},
  numpages = {13},
  year = {2024},
  month = {Apr},
  publisher = {American Physical Society},
  doi = {10.1103/PhysRevA.109.043305},
  url = {https://link.aps.org/doi/10.1103/PhysRevA.109.043305}
}

@article{PhysRevB.110.205402,
  title = {Direct probe of topology and geometry of quantum states on the {IBM Q} quantum processor},
  author = {Chen, Tianqi and Ding, Hai-Tao and Shen, Ruizhe and Zhu, Shi-Liang and Gong, Jiangbin},
  journal = {Phys. Rev. B},
  volume = {110},
  issue = {20},
  pages = {205402},
  numpages = {12},
  year = {2024},
  month = {Nov},
  publisher = {American Physical Society},
  doi = {10.1103/PhysRevB.110.205402},
  url = {https://link.aps.org/doi/10.1103/PhysRevB.110.205402}
}
\end{document}


\lefthyphenmin=2 \righthyphenmin=3
\setcounter{secnumdepth}{1}
\title{Supplemental Material for ``Classical Bound on the Fisher Information Rate of a Dephasing-Enhanced Quantum Neuron''}

\author{Peng Wang}
\email{wp002005@163.com}
\affiliation{School of Computer and Artificial Intelligence, Southwest Minzu University, Chengdu 610225, China}
\affiliation{Centre for Quantum Technologies, National University of Singapore, Singapore 117543, Singapore}

\author{Yu-Xuan Zhang}
\affiliation{School of Physics, Nankai University, Tianjin 300071, China}
\affiliation{Centre for Quantum Technologies, National University of Singapore, Singapore 117543, Singapore}

\author{Hai-Tao Ding}
\affiliation{Centre for Quantum Technologies, National University of Singapore, Singapore 117543, Singapore}
\affiliation{MajuLab, CNRS-UNS-NUS-NTU International Joint Research Unit, Singapore UMI 3654, Singapore}

\author{Leong-Chuan Kwek}
\email{kwekleongchuan@nus.edu.sg}
\affiliation{Centre for Quantum Technologies, National University of Singapore, Singapore 117543, Singapore}
\affiliation{MajuLab, CNRS-UNS-NUS-NTU International Joint Research Unit, Singapore UMI 3654, Singapore}
\affiliation{National Institute of Education, Nanyang Technological University, Singapore 637616, Singapore}

\maketitle

This Supplemental Material gives the derivations supporting the single-qubit results in the Letter. The present analysis builds on our previous work, Ref.~\cite{Zhang2026}, which established the Kraus channel, stationary activation, and strong-dephasing population rates. We reproduce the relevant derivations here to fix notation and make the presentation self-contained. We then derive the gain–bandwidth relation at threshold, the response under repeated measurement, the Fisher information rate of the correlated record, the adaptive-timing bound and its classical saturation, and the finite-dead-time optimization. All parameters are dimensionless except rates and times, and the Poisson attempt rate $\nu$ is held fixed throughout.

\section{Channel construction and exact Bloch equations}

Let $u=\tanh a$, $q=\sech a$, and choose the orthonormal states
\begin{equation}
 \ket{\psi_a}=\sqrt{\frac{1+u}{2}}\ket0+\sqrt{\frac{1-u}{2}}\ket1,
 \qquad
 \ket{\psi_a^\perp}=\sqrt{\frac{1-u}{2}}\ket0-\sqrt{\frac{1+u}{2}}\ket1.
\end{equation}
The target Bloch vector is $\bm n=(q,0,u)$, with $\bm n^2=1$. Define $P=\ket{\psi_a}\bra{\psi_a}$, $Q=\Id-P$, and
\begin{equation}
 A_0=P+sQ,\qquad A_1=\sqrt{1-s^2}\ket{\psi_a}\bra{\psi_a^\perp},\qquad 0\leq s<1.
\end{equation}
Since $PQ=0$ and $P+Q=\Id$,
\begin{equation}
 A_0^\dagger A_0=P+s^2Q,\qquad A_1^\dagger A_1=(1-s^2)Q,
 \qquad \sum_\mu A_\mu^\dagger A_\mu=\Id.
\end{equation}
The Poisson generator is a Lindblad generator~\cite{Lindblad1976} with jump operators $\sqrt\nu A_0$, $\sqrt\nu A_1$, and $\sqrt{\gamma/2}\,Z$. Indeed, using $\mathcal D[J]\rho=J\rho J^\dagger-\{J^\dagger J,\rho\}/2$ gives
\begin{equation}
 \sum_{\mu=0,1}\mathcal D[\sqrt\nu A_\mu]\rho
 +\mathcal D[\sqrt{\gamma/2}\,Z]\rho
 =\nu(\Phi_a\rho-\rho)+\frac\gamma2(Z\rho Z-\rho).
\end{equation}
In particular, no assumption about small $1-s$ is needed.

For $\rho=(\Id+\bm v\cdot\bm\sigma)/2$, the channel acts as
\begin{equation}
 \bm v\longmapsto s\bm v-s(1-s)(\bm n\cdot\bm v)\bm n+(1-s^2)\bm n.
\end{equation}
Writing $\bm v=(c,y,m)$ and $\kappa=\nu(1-s)$ gives
\begin{equation}
 \dot{\bm v}=\kappa\big[(1+s)\bm n-\bm v-s(\bm n\cdot\bm v)\bm n\big]
 -\gamma(c,y,0).
 \label{eq:fullbloch}
\end{equation}
The $c,m$ variables form an autonomous affine system,
\begin{equation}
 \dot{\bm v}_2=\kappa(\bm b-A\bm v_2),\qquad
 \bm v_2=\begin{pmatrix}c\\m\end{pmatrix},\quad
 \bm b=(1+s)\begin{pmatrix}q\\u\end{pmatrix},\quad
 A=\begin{pmatrix}1+sq^2+d&suq\\suq&1+su^2\end{pmatrix},\quad d=\gamma/\kappa.
 \label{eq:A}
\end{equation}
The remaining component satisfies $\dot y=-(\kappa+\gamma)y$. The nonlinearity is a dependence on the externally programmed input $a$. Evolution remains linear in the density operator at every fixed $a$.

\section{Stationary state, physicality, and relaxation}

Let $r=\gamma/[\nu(1-s^2)]$, so $d=(1+s)r$. The determinant of the matrix in Eq.~\eqref{eq:A} is
\begin{equation}
 \det A=(1+s)(1+r+sru^2)>0.
\end{equation}
Solving $A\bm v_{2,*}=\bm b$ yields
\begin{equation}
 c_* =\frac{q}{1+r+sru^2},\qquad
 m_* =\frac{[1+(1+s)r]u}{1+r+sru^2}
      =\frac{(1+\alpha)u}{1+\alpha u^2},\qquad
 \alpha=\frac{sr}{1+r}.
 \label{eq:stationary}
\end{equation}
Thus $\rho_*=(\Id+c_*X+m_*Z)/2$. Complete positivity and convergence already ensure that this is a density operator. An explicit check is useful: if $D=1+r+sru^2$ and $B=1+(1+s)r$, then
\begin{equation}
 D^2-q^2-B^2u^2=(1-u^2)r\,[2+r(1-s^2u^2)]\geq0.
\end{equation}
Consequently $c_*^2+m_*^2\leq1$. At any finite $a$ and $r>0$ the state is mixed; at $r=0$ it is the pure target.

The drift matrix has the positive decomposition
\begin{equation}
 A=I_2+s\begin{pmatrix}q\\u\end{pmatrix}\begin{pmatrix}q&u\end{pmatrix}
 +d\begin{pmatrix}1&0\\0&0\end{pmatrix}\succeq I_2.
\end{equation}
Its eigenvalues are
\begin{equation}
 \lambda_\pm=\frac{2+s+d\pm\sqrt{s^2+d^2+2sd(1-2u^2)}}{2}.
 \label{eq:eigenvalues}
\end{equation}
The Liouvillian eigenvalues are $0,-\kappa\lambda_+,-\kappa\lambda_-,-(\kappa+\gamma)$. All nonzero eigenvalues have real part at most $-\kappa$. The steady state is unique, and deviations decay at least as $e^{-\kappa t}$ in Bloch-vector norm. At $a=0$ the three positive decay rates are $\eta=\nu(1-s^2)+\gamma$, $\kappa$, and $\kappa+\gamma$; the slowest is exactly $\kappa$. A degeneracy of this real symmetric drift matrix is diagonalizable and is not an exceptional point.

Several limits must be distinguished. For $\gamma=0$, $m_*=u$ and $c_*=q$: the output is sigmoid but the state is coherent. For $s=0$, $m_*=u$ and $c_*=q/(1+\gamma/\nu)$, so the populations are noise independent but coherence generally survives. For $r\to\infty$ at fixed $s$, $c_*\to0$ and $m_*\to(1+s)u/(1+su^2)$. A diagonal state with the original sigmoid therefore arises, for example, when $s=0$ and $r\to\infty$, or by discarding coherence after an independent stationary preparation. It does not follow from $\alpha\to0$ alone. Finally, $s=1$ is excluded: the programmed update becomes the identity and no unique population steady state is prepared. This explains why the stationary $s\to1$ gain and the physical relaxation time must be considered together.

\section{Frequency response and stationary information}

For $a(t)=a_0+\delta a(t)$, differentiate the input-dependent matrix and source in Eq.~\eqref{eq:A}. At a fixed operating point the linearized equation is
\begin{equation}
 \delta\dot{\bm v}_2=-\kappa A\delta\bm v_2
 +\kappa(\bm b'-A'\bm v_{2,*})\delta a
 =-\kappa A\delta\bm v_2+\kappa A\bm v_{2,*}'\delta a.
\end{equation}
Primes denote input derivatives at fixed $s,\nu,\gamma$. With an $e^{-i\omega t}$ convention,
\begin{equation}
 \chi(a_0,\omega)=\bm e_m^{\mathsf T}(\kappa A-i\omega I_2)^{-1}
 \kappa A\bm v_{2,*}',\qquad \bm e_m=(0,1)^{\mathsf T}.
\end{equation}
At $a_0=0$, $q'=0$, $u'=1$, and $\bm v_{2,*}'=(0,G)^{\mathsf T}$. Since $A$ is diagonal there,
\begin{equation}
 \chi(0,\omega)=\frac{\kappa G}{\kappa-i\omega},\qquad
 G=1+s\left(1-\frac{1}{1+r}\right).
\end{equation}
This susceptibility describes ensembles under a small input modulation before repeated projective readout. It must not be identified with the transfer function of the measured chain.

The stationary $Z$-readout information has the explicit expression
\begin{equation}
 \mathcal F_Z(a)=\frac{(1+\alpha)^2(1-u^2)(1-\alpha u^2)^2}
 {(1+\alpha u^2)^2(1-\alpha^2u^2)}.
\end{equation}
At the threshold, $\rho_*=(\Id+c_0X)/2$ and $\partial_a\rho_*=GZ/2$. The symmetric logarithmic derivative can be chosen as $L_a=GZ$, because $\{\rho_*,L_a\}/2=\partial_a\rho_*$. The quantum Fisher information~\cite{Braunstein1994} is therefore
\begin{equation}
 \mathcal F_Q(0)=\Tr(\rho_*L_a^2)=G^2=\mathcal F_Z(0).
\end{equation}
This also holds in the rank-one limit $\gamma=0$; it does not require division by the purity deficit. An increase of stationary parameter information does not violate data processing, because the dissipative generator itself depends on the input parameter $a$. No parameter-independent channel acting only after encoding is being compared.

\section{Exact finite-interval measurement process}

After a projective $Z$ outcome $z$, the density operator is $(\Id+zZ)/2$. It is independent of $a$ conditional on the outcome. The $c,m$ propagator for a constant input is
\begin{equation}
 \bm v_2(t)=\bm v_{2,*}+R_a(t)[\bm v_2(0)-\bm v_{2,*}],\qquad
 R_a(t)=e^{-\kappa A(a)t}.
\end{equation}
Define
\begin{equation}
 U_a(\Delta)=[R_a(\Delta)]_{mm},\qquad
 D_a(\Delta)=\bm e_m^{\mathsf T}[I_2-R_a(\Delta)]\bm v_{2,*}.
\end{equation}
The exact transition probabilities, valid at arbitrary finite input, are
\begin{equation}
 p(z'|z;a)=\frac{1+z'[U_a(\Delta)z+D_a(\Delta)]}{2}.
 \label{eq:exacttransition}
\end{equation}
The outcome sequence is a two-state Markov chain because each measurement fixes the entire conditional qubit state. Its stationary mean is
\begin{equation}
 \overline m_\Delta(a)=\frac{D_a(\Delta)}{1-U_a(\Delta)}.
 \label{eq:measuredmean}
\end{equation}
In general, this differs from $m_*(a)$; the distinction concerns the unconditional measured chain as well as its conditional trajectories.

The same transition law gives an exact off-threshold information rate without assuming independent readouts. Denote input derivatives by primes. In the stationary measured chain, the probability of the previous sign is $\pi_z=(1+z\overline m_\Delta)/2$, and the conditional Bernoulli mean of the next sign is $M_z=D_a+zU_a$. Therefore
\begin{equation}
 \mathcal R(a,\Delta)=\frac{1}{\Delta}
 \sum_{z=\pm1}\frac{1+z\overline m_\Delta(a)}{2}
 \frac{[D_a'(\Delta)+zU_a'(\Delta)]^2}
 {1-[D_a(\Delta)+zU_a(\Delta)]^2}.
 \label{eq:generalrate}
\end{equation}
This is the general-$a$ counterpart of the threshold formula below. The channel symmetry gives $U_{-a}=U_a$ and $D_{-a}=-D_a$, so Eq.~\eqref{eq:generalrate} is even in $a$. These facts alone do not prove that $a=0$ is its global maximum, and no such unproved assertion is used in the Letter.

At $a=0$, $U_0=E=e^{-\kappa\Delta}$ and $D_0=0$. During an interval starting from $z$, the zero-input dynamics is $m^{(0)}(t)=ze^{-\kappa t}$ and $c^{(0)}(t)=c_0(1-e^{-\eta t})$. If $h(t)=\partial_a m(t)|_{a=0}$, the longitudinal equation gives
\begin{equation}
 \dot h=-\kappa h+\kappa[(1+s)-sc^{(0)}(t)],\qquad h(0)=0.
\end{equation}
Any first-order coherence generated by the nonzero initial $z$ is multiplied by $u=O(a)$ in the longitudinal feedback, and thus does not enter this first-order equation. Its solution is
\begin{align}
 h_\Delta
 &=\kappa\int_0^\Delta e^{-\kappa(\Delta-t)}
       [(1+s)-sc_0(1-e^{-\eta t})]\,dt\nonumber\\
 &=G(1-e^{-\kappa\Delta})+
 \kappa sc_0\frac{e^{-\kappa\Delta}-e^{-\eta\Delta}}{\eta-\kappa}.
 \label{eq:hsm}
\end{align}
Here $\eta\geq\kappa$; equality occurs only for $s=\gamma=0$. The divided difference is then $\Delta e^{-\kappa\Delta}$ and its prefactor $s$ vanishes. Equation~\eqref{eq:hsm} is independent of the previous sign $z$, as needed for the simple information formula.

By differentiation of Eq.~\eqref{eq:measuredmean},
\begin{equation}
 G_\Delta=\overline m_\Delta'(0)=\frac{h_\Delta}{1-E}.
\end{equation}
Since $0\leq c^{(0)}(t)\leq c_0$, the integral yields $G\leq G_\Delta\leq1+s$. Its limiting values are $G$ at long intervals and $1+s$ at short intervals. At zero input the stationary measured chain is unbiased and has correlation $\langle z_jz_{j+\ell}\rangle=E^{|\ell|}$.

\section{Likelihood information and the adaptive bound}

For a fixed initial sign $z_0$, the record likelihood is
\begin{equation}
 P_a(z_1,\ldots,z_n|z_0)=\prod_{j=1}^{n}p_a(z_j|z_{j-1}),\qquad
 S_n=\partial_a\log P_a=\sum_{j=1}^n\ell_j.
\end{equation}
Here $\ell_j=\partial_a\log p_a(z_j|z_{j-1})$. Normalization implies $\mathbb E_a[\ell_j|z_0,\ldots,z_{j-1}]=0$. Thus $\mathbb E(\ell_i\ell_j)=0$ for $i\ne j$, although the outcomes themselves are correlated. This is the score-increment decomposition used in sequential quantum estimation~\cite{Gammelmark2014,Yang2023,Mentesoglu2026}. At zero input,
\begin{equation}
 \ell_j\big|_{a=0}=\frac{z_jh_\Delta}{1+z_jz_{j-1}E},\qquad
 \mathbb E_0[\ell_j^2|z_{j-1}]=\frac{h_\Delta^2}{1-E^2}.
\end{equation}
No large-$n$ approximation is used. The conditional record information and the rate are
\begin{equation}
 \mathcal F_n=\frac{nh_\Delta^2}{1-E^2},\qquad
 \mathcal R(\Delta)=\frac{\mathcal F_n}{n\Delta}
 =\frac{h_\Delta^2}{\Delta(1-E^2)}
 =G_\Delta^2\frac{\tanh(\kappa\Delta/2)}{\Delta}.
 \label{eq:ratesm}
\end{equation}
For an initially stationary, informative outcome, its initial Fisher information is added. That contribution does not change the long-time rate.

Nonnegativity of $c^{(0)}(t)$ bounds Eq.~\eqref{eq:hsm} by $h_\Delta\leq(1+s)(1-E)$. It follows that
\begin{equation}
 \mathcal R(\Delta)\leq(1+s)^2\frac{\tanh(\kappa\Delta/2)}{\Delta}
 \leq\frac{\kappa(1+s)^2}{2}
 =\frac\nu2(1-s)(1+s)^2\leq\frac{16\nu}{27}.
\end{equation}
The derivative of $(1-s)(1+s)^2/2$ is $(1+s)(1-3s)/2$, so the unique interior maximum is at $s=1/3$. For any fixed finite $\gamma$,
\begin{equation}
 h_\Delta=\kappa(1+s)\Delta+O(\Delta^2),\qquad
 1-E^2=2\kappa\Delta+O(\Delta^2),
\end{equation}
which proves tightness in the limit $\Delta\to0$. Measurement count diverges in this limiting statement, but the time-normalized information remains finite.

For adaptive timing, let $\Delta_j$ be chosen predictably from the already observed history using a rule independent of the unknown $a$. The conditional state after each readout is still a $Z$ eigenstate, and all interval formulas hold with $\Delta\to\Delta_j$. The score remains a martingale difference. For a finite schedule, or an almost surely finite schedule for which the stopped score is square integrable,
\begin{equation}
 \mathcal F_{\mathrm{rec}}
 =\mathbb E_0\sum_j\frac{h_{\Delta_j}^2}{1-e^{-2\kappa\Delta_j}}
 \leq\frac{\kappa(1+s)^2}{2}\mathbb E_0\sum_j\Delta_j.
\end{equation}
For schedules completing at a prescribed total interrogation time $T$, the last expectation equals $T$ and $\mathcal F_{\mathrm{rec}}\leq16\nu T/27$. Parameter-independent randomization of the timing adds no information. This statement permits adaptation of intervals, not adaptation of the input, the channel parameter $s$, or the measurement basis. A separate informative initial preparation contributes an additive initial term.

\section{Finite measurement dead time and joint optimization}

We model detector dead time operationally: the input-sensitive evolution lasts for $\Delta$, a projective outcome is recorded, and the resulting eigenstate is held for a noninformative time $\tau_m$ before the next interval begins. The score information per cycle is still $h_\Delta^2/(1-E^2)$, but the elapsed time is $\Delta+\tau_m$. Thus
\begin{equation}
 \mathcal R_{\tau_m}(\Delta;s,\gamma)
 =\frac{h_\Delta^2}{(\Delta+\tau_m)(1-e^{-2\kappa\Delta})}
 =\mathcal R(\Delta;s,\gamma)\frac{\Delta}{\Delta+\tau_m}.
 \label{eq:deadsm}
\end{equation}
This hold-time convention is essential. If the input-dependent channel continues to evolve unobserved throughout detector recovery, its duration must instead be included in the propagator interval.

For $\tau_m>0$, Eq.~\eqref{eq:deadsm} tends to zero at both $\Delta\downarrow0$ and $\Delta\to\infty$, so dead time regularizes the rapid-readout supremum into a finite-cadence optimum. Set
\begin{equation}
 \delta=\nu\Delta,\qquad \theta=\nu\tau_m,\qquad g=\gamma/\nu.
\end{equation}
The joint optimum over $(s,\delta)$ depends only on $(g,\theta)$ and scales overall as $\nu$. In the zero and strong-dephasing limits, the rate reduces to simple analytic expressions. At zero dephasing,
\begin{equation}
 \frac{\mathcal R_{\tau_m}^{(0)}}{\nu}
 =\frac{[1-e^{-(1-s^2)\delta}]^2}
 {(\delta+\theta)[1-e^{-2(1-s)\delta}]},
 \label{eq:deadzero}
\end{equation}
whereas the strong-dephasing envelope is
\begin{equation}
 \frac{\mathcal R_{\tau_m}^{(\infty)}}{\nu}
 =(1+s)^2\frac{\tanh[(1-s)\delta/2]}{\delta+\theta}.
 \label{eq:deadinf}
\end{equation}

\begin{figure}[!t]
 \includegraphics[width=0.82\textwidth]{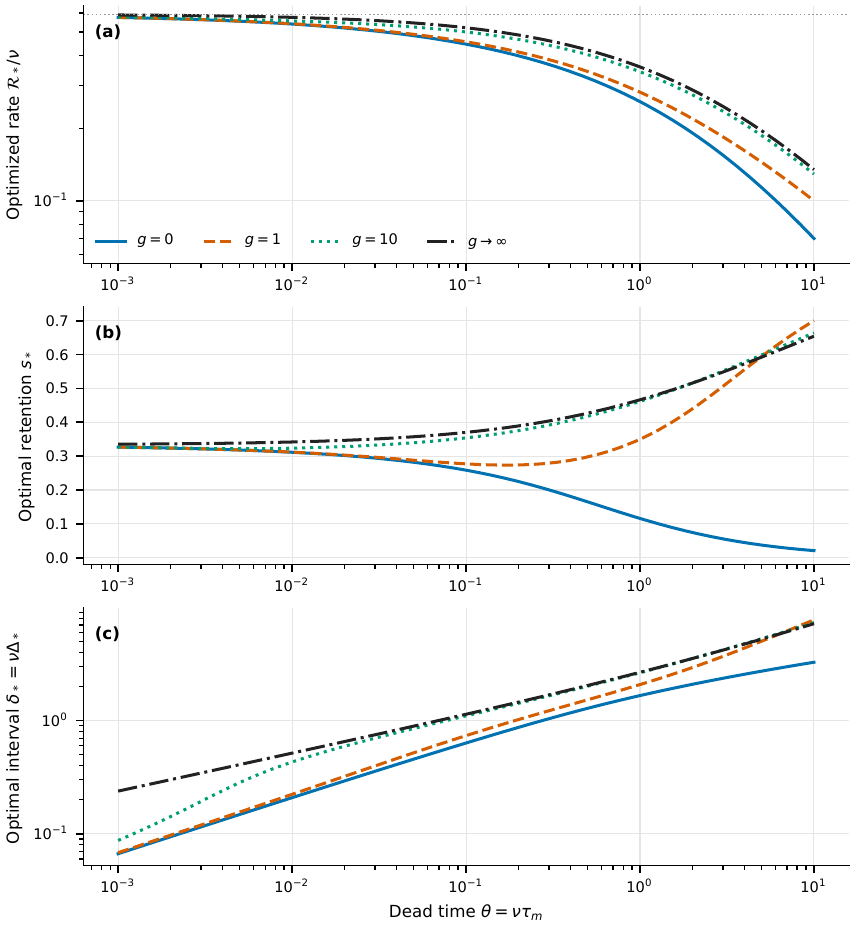}
 \caption{Joint dead-time optimization. (a) Optimized information rate, (b) optimal retention, and (c) optimal free-evolution interval versus dimensionless dead time. Each finite-$g$ curve maximizes Eq.~\eqref{eq:deadsm} over $0\leq s<1$ and $\delta>0$; the black dash-dotted curve is the analytic strong-dephasing envelope in Eq.~\eqref{eq:deadinf}.}
 \label{fig:deadtime}
\end{figure}

Figure~\ref{fig:deadtime} shows the bounded numerical maximization of Eq.~\eqref{eq:deadsm}; representative values are listed in Table~\ref{tab:deadtime}. Even modest dead time moves the best retention away from $1/3$ and the best interval away from zero. Dephasing can then improve the attainable finite-cadence rate, although it never changes the zero-dead-time ceiling in the Letter.

\begin{table}[!t]
\caption{Selected joint optima $(s_*,\delta_*,\mathcal R_*/\nu)$, where $\delta_*=\nu\Delta_*$. The $g=\infty$ rows use Eq.~\eqref{eq:deadinf}.}
\label{tab:deadtime}
\begin{ruledtabular}
\begin{tabular}{ccccc}
$\theta$ & $g$ & $s_*$ & $\delta_*$ & $\mathcal R_*/\nu$\\
\hline
0.01 & 0        & 0.311 & 0.209 & 0.540\\
0.01 & 1        & 0.312 & 0.223 & 0.542\\
0.01 & $\infty$ & 0.342 & 0.516 & 0.576\\
0.10 & 0        & 0.259 & 0.634 & 0.446\\
0.10 & 1        & 0.277 & 0.737 & 0.456\\
0.10 & $\infty$ & 0.370 & 1.138 & 0.521\\
1.00 & 0        & 0.116 & 1.664 & 0.258\\
1.00 & 1        & 0.350 & 2.082 & 0.282\\
1.00 & $\infty$ & 0.467 & 2.671 & 0.359\\
\end{tabular}
\end{ruledtabular}
\end{table}

\section{Classical saturation and resource accounting}

Let $\mathcal P$ discard off-diagonal density-matrix elements in the $Z$ basis. The infinitesimal repeated-readout evolution on populations has generator $\mathcal P\mathcal L_a\mathcal P$. The dephasing term vanishes under this projection. Starting in a diagonal eigenstate of sign $z$, the Bloch equation gives $\dot m=\kappa[(1+s)u-(1+su^2)z]$. Since a flip changes $m$ by $-2z$, the rate is
\begin{equation}
 w(z\to-z)=\frac\kappa2[1+su^2-(1+s)zu]
 =\frac\kappa2(1-zu)(1-szu).
\end{equation}
These coincide with the established strong-dephasing rates~\cite{Zhang2026}. At $a=0$, both rates equal $\kappa/2$, and
\begin{equation}
 \partial_a\log w(z\to-z)\big|_{a=0}=-(1+s)z.
\end{equation}
The continuous-time trajectory information is the expected jump intensity times the squared logarithmic derivative, giving $\mathcal R_{\mathrm{cl}}=\kappa(1+s)^2/2$. The classical process attains the information bound; no coherence-sensitive measurement is needed at the optimum. This identifies the optimized ceiling as a classical optimal ceiling for the projected channel, even though finite-time quantum coherence affects nonoptimal cadences. Frequent projection suppresses coherence but not the first-order population jump rates of the assumed Markovian generator. Replacing that generator by a microscopic finite-memory evolution could change the very-short-time limit.

Only $\nu$ is fixed in this optimization. The polarization relaxation scale $\nu(1-s^2)$ and the threshold bandwidth $\nu(1-s)$ vary with $s$. The factor $32/27$ compares the optimum with $s=0$ at the same $\nu$, using the same input convention $u=\tanh a$ and ideal measurement accounting. It is not a bound for all classical or quantum neurons. If the standard logistic argument is instead $A=2a$, the information about $A$ is one quarter of the information about $a$; the physical conclusion and the optimal $s$ are unchanged. Finite readout precision, uncertainty in $\gamma$, joint parameter estimation, and information extracted from the environment remain outside the tightness claim.

\bibliography{references}